\documentclass[a4paper,11pt]{article}
\pdfoutput=1 

\usepackage{jheppub}
\usepackage{epsfig}
\usepackage{amsmath}
\usepackage{graphicx}

\def\A0#1{\Pi_{\rm #1}(0)}
\def\AP0#1{\Pi'_{\rm #1}(0)}
\def\be{\begin{equation}}
\def\ee{\end{equation}}
\def\bea{\begin{array}}
\def\eea{\end{array}}
\def\beqa{\begin{eqnarray}}
\def\eeqa{\end{eqnarray}}
\def\beqas{\begin{eqnarray*}}
\def\eeqas{\end{eqnarray*}}

\def\bp{\begin{picture}}
\def\ep{\end{picture}}
\def\bc{\begin{center}}
\def\ec{\end{center}}
\def\bfig{\begin{figure}}
\def\efig{\end{figure}}

\def\bit{\begin{itemize}}
\def\eit{\end{itemize}}
\def\nn{\nonumber}
\def\f{\frac}

\def\[{\left[}
\def\]{\right]}
\def\({\left(}
\def\){\right)}

\def\..{\left.}
\def\.{\right.}

\def\ra{\rightarrow}

\def\tm{\times}

\def\da{\dagger}

\def\al{\alpha}

\def\ka{\kappa}

\def\ep{\epsilon}

\def\ga{\gamma}

\title{Explaining The CDF-II W-Boson Mass Anomaly in the Georgi-Machacek Extension Models}
\author[a]{Xiao Kang Du,}
\author[a]{Zhuang Li,}
\author[a]{Fei Wang,}
\author[a]{Ying Kai Zhang,}

\affiliation[a]{School of Physics, Zhengzhou University, Zhengzhou 450000, P. R. China}
\emailAdd{feiwang@zzu.edu.cn}
\abstract{ Original Georgi-Machacek model can preserve the custodial symmetry at tree level
after the electroweak symmetry breaking. Unless additional $SU(2)_c$ custodial symmetry breaking effects are significant, the new physics contributions to $\Delta m_W$ are always very small. Our numerical results show that ordinary GM model can contribute to $\Delta m_W$ a maximal amount $0.0012$ GeV, which can not explain the new CDF-II anomaly on W boson mass.  We propose firstly to introduce small misalignment among the triplet VEVs to increase $\Delta m_W$, which can give large new physics contributions to $\Delta m_W$. Such slightly misaligned triplet VEVs from custodial symmetry preserving scalar potential can still be allowed. Our numerical results indicate that the resulting $\Delta m_W$ can easily reach the $1\sigma$ range of CDF-II $m_W$ data and the splitting among the triplet VEVs $\Delta v$($\equiv v_\xi-v_\chi$) can be as small as $0.8$ GeV for $v_\chi\lesssim 15$ GeV. We also propose to introduce an additional custodial symmetry breaking source by extending the GM model with a low scale RH neutrino sector, which can adopt the leptogenesis mechanism and allow moderately large coupling strength $h_{ij}$ even for triplet VEVs of order GeV. With low scale RH neutrino mass scale of order $10^2\sim 10^4$ TeV, the new physics contribution to $\Delta m_W$ can reach $0.03$ GeV and is much larger than that of ordinary GM model. Combining both custodial $SU(2)_c$ symmetry breaking effects, the small misalignment among the triplet VEVs and the moderately large $h_{ij}$ couplings allowed with RH neutrino sector, the value of $\Delta m_W$ can still easily reach the $1\sigma$ range of CDF-II $m_W$ data, with a minimum splitting (among the triplet VEVs) approximately $0.7$ GeV for $v_\chi\lesssim 15$ GeV.}
\begin{document}
\maketitle
\newpage
\section{Introduction}
The standard model (SM) had already been corroborated by various contemporary collider experiments, including the discovery of 125 GeV SM-like Higgs boson by the Large Hadron Collider (LHC)~\cite{ATLAS:higgs,CMS:higgs}. Despite its immense success, it has many theoretical or aesthetical problems, such as the hierarchy problem related to the fundamental Higgs scalar, the origin of tiny neutrino masses, the origin of the baryon asymmetry in the universe and the measured W boson mass data recently reported by CDF-II experiment at the Fermilab~\cite{CDF:W}. In fact, the CDF-II experiment reported recently the most precise direct measurement of the W boson mass as
\beqa
m_W=80,433.5 \pm  6.4(stat) \pm 6.9 (syst)=80,433.5\pm 9.4 MeV/c^2~,
\eeqa
using data corresponding to $8.8 fb^{-1}$ of integrated luminosity collected in proton-antiproton collisions at a 1.96 TeV center-of-mass energy. This measurement is in significant tension with the SM expectation~\cite{SM:W}
\beqa
m_W=80,357\pm 4(inputs)\pm 4 (theory) MeV/c^2~.
\eeqa
The data-driven techniques used by CDF experiment capture most of the higher order corrections and using higher order corrections result in a decrease in the W boson mass value reported by CDF-II by almost 10 MeV~\cite{Isaacson:2022rts}. So, the deviation of CDF-II $m_W$ value from the SM expectation can be reduced at most from $7\sigma$ to $6\sigma$, necessitating additional theoretical or experimental explanations.

Although the newly reported CDF-II measurement of $m_W$ is still $2.4 \sigma$ away from the combined results of ATLAS~\cite{ATLAS:W}, LHCb~\cite{LHCb:W} and LEP, such a discrepancy may be a hint for the new physics beyond the SM~\cite{anomaly:W}, especially if such an anomaly persists and gets confirmed by other experiments. Additional contributions to $m_W$ may need TeV scale new particles that transform non-trivially under $SU(2)_L$ or tiny mixing between the ordinary W gauge boson and some heavy new gauge bosons etc, see various explanations in~\cite{W1,W3,W4,W5,W6,W7,W8,W9,W10,W11,W12,W13,W14,W15,W16,W17,W18,W19,W20,W21,W22,W23,W24,W25,W26,Cao:2022mif}. We should note that, the CDF-II measurement of $m_W$ should better be seen as an additional constraint for predictive new physics models, for example, models with extended Higgs sector. 

 Electroweak symmetry breaking (EWSB) mechanism with extended Higgs sector is still allowed, given the uncertainties in Higgs boson coupling measurements. On the other hand, any extended Higgs sector must be carefully constructed to satisfy the stringent constraints from electroweak precision measurments, the most important one of which is the electroweak $\rho$ parameter. We know that tree level relation $\rho_{tree}=1$ is automatically satisfied by the Higgs sector of SM, which respects the custodial $SU(2)_c$ global symmetry. Georgi-Machacek (GM) model~\cite{GM,GM2,GM3,GM4,GM5,GM6,GM7,GM8,GM9,GM10,GM11,GM12,GM13,GM14,GM15,GM16,GM17,
 GM18,GM19,GM20,GM21,GM22,GM23,GM24,GM24v2,GM25,GM26,GM27,GM28,GM28.5,GM29,GM30,GM31}, which augments the SM Higgs sector with a complex $SU(2)_L$ triplet of hypercharge $Y=1$ and a real $SU(2)_L$ triplet of $Y=0$, can protect the relation $\rho_{tree}=1$ with custodial symmetry preserving Higgs potential and vacuum alignment between the complex and real triplets. Therefore, a large triplet VEV $v_\Delta$ of order ${\cal O}(10)$ GeV is still allowed for the vacuum aligned triplets in the GM model, unlike that in the minimal Higgs triplet model (HTM) whose $v_\Delta$ is constrained to be much smaller because of its violation of $SU(2)_c$ at tree level. GM model can adopt the type-II neutrino seesaw mechanism to generate tiny neutrino masses.

 However, we expect that the new physics contribution $\Delta m_W$ in GM model should be small because of the tree-level custodial $SU(2)_c$ symmetry, which protects the $\rho$ parameter from large deviations from unity. So, in order to increase $\Delta m_W$, relatively large custodial $SU(2)_c$ breaking effects should be introduced.  We propose firstly to spoil the $SU(2)_c$ custodial symmetry with slightly misalignment among the two triplet VEVs, which can be natural because the global symmetry of the tree level scalar potential, after weakly gauged, is not fully respected by the 1-loop effective scalar potential. We also propose an alternative GM extension model, which augments the neutrino sector with low scale right-handed (RH) neutrinos and adopts the type I+II seesaw like mechanism similar to that in the Left-Right symmetric model. The presence of RH neutrino terms can not only be used to increase the value of  neutrino-scalar couplings $h_{ij}$ and the masses of the light pseudo-Goldstone modes from $SU(2)_c$ to residual $U(1)$ breaking, but also be used to understand the Baryon Asymmetry of the Universe (BAU). Satisfying Sakharov's three conditions, leptogenesis mechanism naturally combine the neutrino seesaw mechanism and net baryon number generation by out-of-equilibrium L-violating decays of heavy Majorana neutrinos. Besides, low scale RH neutrino with large Yukawa couplings to leptons can have rich collider phenomenology.

This paper is organized as follows. In Sec~\ref{sec-2}, we review the GM model and the GM extension model with slightly misaligned triplet VEVs. In Sec~\ref{sec-3}, we show our numerical results on the new physics contributions to $\Delta m_W$ in custodial preserving GM model and extended GM model with slightly misaligned triplet VEVs, respectively. In Sec~\ref{sec-4}, we discuss the low scale RH neutrino extended GM models.
Numerical results of $\Delta m_W$ are given for the cases with (and without) slightly misaligned triplet VEVs, respectively. Sec~\ref{sec-5} contains our conclusions.

\section{\label{sec-2} GM model and its extension with slightly misaligned triplet VEVs}
In the GM model~\cite{GM,GM13}, the Higgs sector contains the ordinary SM $SU(2)_L$ doublet Higgs field $\phi$ with hypercharge $Y=1/2$ and two $SU(2)_L$ triplet Higgs fields: complex triplet Higgs $\chi$ with $Y=1$ and real triplet $\xi$ with $Y=0$. These fields can be written in the form of $SU(2)_L\times SU(2)_R$ symmetry
\begin{align}
\Phi=\left(
\begin{array}{cc}
\phi^{0*} & \phi^+ \\
\phi^- & \phi^0
\end{array}\right),\quad
\Delta=\left(
\begin{array}{ccc}
\chi^{0*} & \xi^+ & \chi^{++} \\
\chi^- & \xi^0 & \chi^{+} \\
\chi^{--} & \xi^- & \chi^{0}
\end{array}\right), \label{eq:Higgs_matrices}
\end{align}
to track the symmetry preserved by the scalar potential, where the transformations of $\Phi$ and $\Delta$ under $SU(2)_L\times SU(2)_R$ are given as
$\Phi\to U_L\Phi U_R^\dagger$ and $\Delta\to U_L\Delta U_R^\dagger$ with
$U_{L,R}=\exp(i\theta_{L,R}^aT^a)$ and $T^a$ being the $SU(2)$ generators.
The global symmetry can be weakly gauged so as that the $SU(2)_L, U(1)_Y$ group can be associated with the $T^a_L$ and $T^3_R$ generators of $SU(2)_L\times SU(2)_R$, respectively.
The gauge invariant scalar potential without $\Delta \ra -\Delta$ discrete $Z_2$ symmetry are given as~\cite{GM13}
\begin{align}
V(\Phi,\Delta) = & \frac12 m_{\Phi}^2 {\rm tr} \left[ \Phi^\dagger \Phi \right] + \frac12 m_{\Delta}^2 {\rm tr} \left[ \Delta^\dagger \Delta \right] + \lambda_1 \left( {\rm tr} \left[ \Phi^\dagger \Phi \right] \right)^2 + \lambda_2 \left( {\rm tr} \left[ \Delta^\dagger \Delta \right] \right)^2 \nonumber \\
& + \lambda_3 {\rm tr} \left[ \left( \Delta^\dagger \Delta \right)^2 \right] + \lambda_4 {\rm tr} \left[ \Phi^\dagger \Phi \right] {\rm tr} \left[ \Delta^\dagger \Delta \right] + \lambda_5 {\rm tr} \left[ \Phi^\dagger \frac{\sigma^a}{2} \Phi \frac{\sigma^b}{2} \right] {\rm tr} \left[ \Delta^\dagger T^a \Delta T^b \right] \nonumber \\
& + \mu_1 {\rm tr} \left[ \Phi^\dagger \frac{\sigma^a}{2} \Phi \frac{\sigma^b}{2} \right] (P^\dagger \Delta P)_{ab} + \mu_2 {\rm tr} \left[ \Delta^\dagger T^a \Delta T^b \right] (P^\dagger \Delta P)_{ab}
~,
\label{eq:GMpot}
\end{align}
where $\sigma^a$ are the Pauli matrices, $T^a$ are the $3\times3$ matrix representation of the $SU(2)$ generators, and the similarity transformation, which rotates $\Delta$ into the Cartesian basis, is given by
\begin{align*}
P = & \frac{1}{\sqrt{2}} \left( \begin{array}{ccc} -1 & i & 0 \\ 0 & 0 & \sqrt{2} \\ 1 & i & 0 \end{array} \right).
\end{align*}
The scalar potential can trigger successful EWSB with the scalar VEVs
\beqa
\langle \phi\rangle=\f{1}{\sqrt{2}} v_\phi~,~~~\langle \chi\rangle=v_\chi~,
~~~\langle \xi\rangle= v_\xi~.
\eeqa

When the two triplet VEVs align, $v_\chi=v_\xi \equiv v_\Delta$, the custodial $SU(2)_c$ symmetry from the diagonal breaking of the global $SU(2)_L\times SU(2)_R$ symmetry of the potential~\footnote{Global $SU(2)_L\times SU(2)_R$ is explicitly broken by the Yukawa and the hypercharge gauge interactions.} is preserved.
The EWSB condition
\beqa
v^2=v_\phi^2+4v_\chi^2+4v_\xi^2=v_\phi^2+8v_\Delta^2=\f{1}{\sqrt{2}G_F}\approx \(246 {\rm GeV}\)^2~,
\eeqa
can be recasted into $s_H$ variable with $s_H=\sin\theta_H$ for
\beqa
\tan\theta_H=2\sqrt{2}v_\Delta/v_\phi~.
\eeqa
The gauge boson masses at tree level can be given as
\beqa
m_W=\f{g_2^2}{4}\(v_\phi^2+4v_\chi^2+4v_\xi^2\)~,~m_Z^2=\f{g_2^2}{4\cos^2\theta_W}\(v_\phi^2+8v_\chi^2\)~,
\eeqa
which obviously predicts $\rho_{tree}\equiv m_W^2/M_Z^2\cos^2\theta_W=1$  when $v_\chi=v_\xi \equiv v_\Delta$. In fact, in multiple Higgs scenarios, the tree level $\rho$ parameter can be expressed as the sum
of contributions from Higgs multiplets $\phi_{T,Y}$ with the corresponding hypercharge $Y\equiv 2(Q-T_3)$, isospin $T$ and the VEV $\langle\phi_{T,Y}\rangle=V_{T,Y}$~\cite{gunion}
\beqa
\rho_{tree}\equiv\f{m_W^2}{m_Z^2\cos^2\theta_W}=\f{\sum\limits_{T,Y}\[4T(T+1)-Y^2\]|V_{T,Y}|^2 c_{T,Y}}
{2\sum\limits_{T,Y}Y^2|V_{T,Y}|^2}~,
\eeqa
for
\beqa
c_{T,Y}=\left\{\bea{l}~1,~~(T,Y)\in {\rm complex~representation,} \\~\f{1}{2},~~(T,Y=0)\in {\rm real~ representation.} \eea \right.
\eeqa

After introducing the Yukawa interactions between the lepton doublets and the Higgs triplet
\beqa
{\cal L}\supseteq h_{ij}\overline{L_L^{ic}}i\tau_2\chi L_L^j+\text{h.c.}~,
\label{neutrino:typeII}
\eeqa
tiny Majorana neutrino masses can be generated
\beqa
m_\nu\approx h_{ij}v_\Delta~.
\eeqa
For $v_\Delta\sim {\cal O}(10)$ GeV, very small coupling $h_{ij}\sim 10^{-13}$ is needed to give tiny neutrino mass, which is rather unnatural.

The scalar potential with $\Delta \ra -\Delta$ symmetry, originally proposed in~\cite{GM2}, can naturally eliminate the $\mu_1,\mu_2$ cubic terms in the Higgs potential~(\ref{eq:GMpot}) and lead to the custodial symmetry preserving minimum with $v_\chi=v_\xi \equiv v_\Delta$. In fact, such a $Z_2$ symmetric potential can  not only preserves the $\rho=1$ condition at tree-level but also offers the possibility that the triplets give large contribution to the EWSB. Detailed discussions on the vacuum structure of this $Z_2$ symmetric scalar potential are given in~\cite{GM31}. On the other hand, it had been shown that the custodial  symmetry preserving minimum may not be safe from eventual tunneling to a deeper non-custodial vacuum or a charge breaking vacuum. In fact, it was shown in~\cite{GM20} that a sufficient (but not necessary) condition for the custodial symmetry preserving vacuum is given by $\lambda_3>0,\lambda_5<0, \mu_1<0,\mu_2<0$. Choices of inputs to generate non-custodial symmetry preserving vacuum from $SU(2)_L\tm SU(2)_R$ symmetry preserving scalar potential can still be possible. Explicit expressions for the difference of the scalar potential values between real non-custodial vacuum $ V_{NC}$ and the custodial vacuum $V_{C}$ in the $Z_2$ symmetric case had been calculated in~\cite{GM31}, which can take either signs with proper choices of parameters. Adding the mass parameters $\mu_1,\mu_2$ related terms will enlarge the allowed parameters space for real non-custodial minimum.  Therefore, it is possible that the true vacuum does not keep the custodial symmetry and takes only the real non-custodial one. In addition, the tree level scalar potential will receive loop corrections from custodial symmetry breaking hypercharge gauge and Yukawa interactions. The one-loop effective potential includes the tree-level scalar potential and the Coleman-Weinberg potential~\cite{CW} is given by
\begin{align}
V_{CW}^1(\varphi, \, \mu_R)
=& \frac{1}{64\pi^2} \left[
\sum_{i } n_i \, (m_i^2(\varphi))^2 \, \left\{ \log\left( \frac{m_i^2(\varphi)}{\mu_R^2} \right) - C_i \right\} \right],
\label{zeroeff}
\end{align}
where $m_i^2(\varphi)$ denotes field-dependent mass squared of mass eigenstate species $i$ (with $i$ runs over particle species), $n_i$ counts degrees of freedom (with a minus sign for fermions), $\mu_R$ is the renormalization scale and $C_i$'s are constants that depend on the renormalization scheme. The particle species include the $W$ and $Z$ bosons, the (field-dependent) mass eigenstates of Higgs bosons, the would-be Nambu-Goldstone modes and the top quarks (light fermions with small Yukawa couplings can be neglected ). See~\cite{GM21} for discussions and applications in GM model.
Taking into such non-custodial symmetry preserving loop contributions, it is also natural for the one-loop effective potential to choose the real non-custodial minimum and slightly misaligned $v_\chi$ and $v_\xi$. Such custodial symmetry breaking vacuum with misaligned triplet VEVs is welcome to explain the new CDF-II data on $m_W$. Substitution the misaligned VEVs for triplets and the doublet VEVs
\begin{align}
\Phi=\left(
\begin{array}{cc}
\phi^{0*}+\f{1}{\sqrt{2}}v_\phi & \phi^+ \\
\phi^- & \phi^0+\f{1}{\sqrt{2}}v_\phi
\end{array}\right),\quad \Delta=\left(
\begin{array}{ccc}
\chi^{0*}+ v_\chi & \xi^+ & \chi^{++} \\
\chi^- & \xi^0+ v_\xi & \chi^{+} \\
\chi^{--} & \xi^- & \chi^{0}+v_\chi
\end{array}\right), \label{eq:Higgs_matrices}
\end{align}
we can obtain the low energy scalar potential, which preserves residual $U(1)$ instead of the custodial $SU(2)$ symmetry.

The most general $SU(2)_L \tm U(1)_Y$ invariant tree-level scalar potential is given by~\cite{GM28.5}
\begin{align}
V(\phi,\chi,\xi)&=m_\phi^2(\phi^\dagger \phi)+m_\chi^2\text{tr}(\chi^\dagger\chi)+m_\xi^2\text{tr}(\xi^2)\notag\\
&+\mu_1\phi^\dagger \xi\phi +\mu_2 [\phi^T(i\tau_2) \chi^\dagger \phi+\text{h.c.}] +\mu_3\text{tr}(\chi^\dagger \chi \xi)+\lambda (\phi^\dagger \phi)^2 \notag\\
&
+\rho_1[\text{tr}(\chi^\dagger\chi)]^2+\rho_2\text{tr}(\chi^\dagger \chi\chi^\dagger \chi)
+\rho_3\text{tr}(\xi^4)
+\rho_4 \text{tr}(\chi^\dagger\chi)\text{tr}(\xi^2)
+\rho_5\text{tr}(\chi^\dagger \xi)\text{tr}(\xi \chi)\notag\\
&+\sigma_1\text{tr}(\chi^\dagger \chi)\phi^\dagger \phi+\sigma_2 \phi^\dagger \chi\chi^\dagger \phi
+\sigma_3\text{tr}(\xi^2)(\phi^\dagger \phi)
+\sigma_4 (\phi^\dagger \chi\xi \phi^c + \text{h.c.}), \label{pot_gen}
\end{align}
where $\phi^c=i\tau_2\phi^*$. It was pointed out in~\cite{1804.02633} that one can introduce custodial symmetry breaking terms from the beginning to make the model consistent at loop levels. They noted that the minimal extension of the Higgs potential is to introduce explicitly custodial symmetry breaking mass term $m_\xi^2\xi^\da\xi$ in addition to ordinary custodial preserving GM scalar potential, which acts as a special case of~(\ref{pot_gen}) and is also adopted in~\cite{2204.12898}. Such a choice can potentially avoid undesirable light masses for pseudo-Goldstone modes $H_5^\pm$ and spoil the bad mass relations $m^2_{H_5^{\pm\pm}}\simeq -3m^2_{H_5^0}$~\footnote{This mass relation will be changed by loop corrections to the scalar masses in a UV completed theory.}, which is observed in a large portion of parameter space and can potentially cause the breaking of the $U(1)_{EM}$ symmetry~\cite{2204.12898}. On the other hand, no disruptive jumping transition from the true stable vacuum in the aligned case to an unstable saddle point vacuum in the misaligned case should occur when one begins to slightly turn on small misalignment among the triplet VEVs (unless the splitting within the misaligned triplet VEVs is big enough) from a custodial symmetry preserving tree level scalar potential (appended with custodial symmetry breaking loop level effective potential) by continuously varying the relevant parameters.
So, it is reasonable that small viable parameter space can be found to get a stable vacuum (or a metastable one with long enough lifetime) with small splitting among the triplet VEVs from the custodial symmetry preserving scalar potential  so as that such a vacuum is not bothered with the previously mentioned extra-light pseudo-Goldstone $H_5^\pm$ masses and negative $m^2_{H_5^{\pm\pm}}$ problems. So, our case with small (but non-vanishing) splitting among the misaligned triplet VEVs corresponds to a stable vacuum (or a metastable one with long enough lifetime), which connects smoothly to the aligned true stable vacuum before its disruptive transition by jump to the vacuum that bothered with the previous noted problems when the splitting among the misaligned triplet VEVs is too large.

The breaking of $SU(2)_L\tm SU(2)_R$ global symmetry into residual $U(1)$ instead of custodial $SU(2)_c$ will lead to two additional light pseudo-Goldstone modes, which can be identified to be $H_5^\pm$ and face various phenomenological constraints. Such pseudo-Goldstone modes can be compatible with current collider exclusion bounds because the global symmetries $SU(2)_L\tm SU(2)_R$ of the tree level potential are not fully respected by the electroweak gauge group and Yukawa couplings. Taking into account the loop corrections to the scalar potential, it is reasonable that the custodial symmetry breaking 1-loop effective scalar potential can give additional contributions to the masses of the pseudo-Goldstone modes. On the other hand,  additional custodial $SU(2)_c$ breaking sources are still welcome to increase further the masses of the otherwise Goldstone modes so as that current collider constraints can be satisfied safely. One can simply introduce the soft custodial breaking triplet mass terms in the scalar potential so as the generalized GM model scalar potential is used~\cite{GM28.5}. Contributions from moderately large Yukawa couplings involving the neutrinos and $\chi$ triplet can also be adopted to push heavy such pseudo-Goldstone modes, which can potentially be advantageous to explain the CDF-II data on W boson mass in the GM framework. Moderately large Yukawa couplings for neutrinos can be well motivated when the RH neutrino sector is introduced to adopt a mixed type I+II like neutrino seesaw mechanism. The pseudo-Goldstone modes $H_5^\pm$ can receive quadratic divergence contributions from scalar, gauge loops and possible lepton loops from the neutrino Yukawa coupling term~(\ref{neutrino:typeII}). Although additional mechanism from UV completion can be introduced to tame the quadratic divergence of the scalar self energy, the scalar masses can still receive large contributions from new heavy particles in the UV completed theory, for example, the sneutrino loops in the SUSY framework. In general, with proper UV completion, the new pseudo-Goldstone modes $H_5^\pm$ from the breaking of custodial $SU(2)_c$ into residual $U(1)$ can be heavy enough to survive the collider exclusion bounds. Especially, when the coupling strengths $h_{ij}$ are large, the UV completion scale can be as low as the TeV scale, which is still large enough to push heavy the possible Goldstone modes so as the the collider bounds can be passed.

\section{\label{sec-3} Contributions to $\Delta m_W$ with slightly misaligned triplet VEVs}
If the mass scale of new physics is much higher than $m_Z$, it contributes to the electroweak precision observables only through virtual loops. The dominant loop contributions to the EW precision observables can be parameterized by the oblique parameters $S, T$ and $U$, which characterize the influence of physics beyond the standard model on
the experimentally measurable quantities in terms of their contributions to the usual gauge boson propagators. They can also be seen as the reparameterizations of the variables $\Delta \rho$, $\Delta\ka$ and $\Delta r$, which absorb the radiative corrections into the quantities that characterize the deviations from the tree-level value of $\rho$, the weak mixing angle and the Fermi constant, respectively.

The W boson mass in the SM can be precisely calculated with the following set of input parameters: \beqas G_F,~ m_Z,~m_t,~m_h,~\alpha_{em}(0),~\alpha_s(M_Z)~.\eeqas
Using the central values of input parameters, the calculated W boson mass $m_W = 80.357 {\rm GeV}$ is $7\sigma$ away from the CDF-II data, necessitating additional contributions $\Delta m_W$ from new physics. The oblique parameters can be used to calculate the loop contributions of new physics to the W-boson masses~\cite{STU,STU1,STU2,W:STU}
\beqa
\Delta m_W=\f{\al M_W}{2(c_W^2-s_W^2)}\(-\f{1}{2}S+c_W^2 T+\f{c_W^2-s_W^2}{4s_W^2} U\)~,
\eeqa
with
\beqa
\alpha S & = & 4s_W^2 c_W^2
               \left[ \AP0{ZZ}
                          -\frac{c_W^2-s_W^2}{s_W c_W}\AP0{Z\gamma}
                          -\AP0{\gamma\gamma}
               \right]\,,  \nonumber \\
\alpha T & = & \frac{\A0{WW}}{m_W^2} - \frac{\A0{ZZ}}{m_Z^2}\,, \\
\alpha U & = & 4s_W^2
               \left[ \AP0{WW} - c_W^2\AP0{ZZ}
                         - 2s_Wc_W\AP0{Z\gamma} - s_W^2\AP0{\gamma\gamma}
               \right]\,, \nonumber
\eeqa
 in additional to possible tree-level shift of $m_W$ as $\Delta m_W^{tree} \approx m_W^{SM} {\Delta \rho}/{2}$, which will be present if the tree level custodial $SU(2)_c$ symmetry is violated.
In our cases, we have
 \beqa
\Delta\rho&\equiv&\rho-1=\f{v_\phi^2+4v_\chi^2+4v_\xi^2}{v_\phi^2+8v_\chi^2}-1= \f{v_{EW}^2}{v_{EW}^2+4(v_\chi^2-v_\xi^2)}-1\approx \f{4(v_\xi^2-v_\chi^2)}{v_{EW}^2}~,\eeqa
with $v_{EW}^2=v_\phi^2+4v_\chi^2+4v_\xi^2\approx(246 {\rm GeV})^2$.
So, if the CDF-II $m_W$ anomaly is explained to $1\sigma$ range by pure tree-level custodial symmetry breaking contributions
\beqa
\f{1}{2}\f{4(v_\xi^2-v_\chi^2)}{v_{EW}^2}m_W\gtrsim 0.0671 {\rm GeV}~,
\eeqa
the misaligned triplet VEVs need to satisfy
\beqa
v_\xi^2-v_\chi^2 \gtrsim 25.27 {\rm (GeV)^2}.
\eeqa
Taking into account the loop contributions, such a bound for misaligned triplet VEVs can be slightly weaken.

\begin{figure}[htb]
\begin{center}
\includegraphics[width=3.5 in]{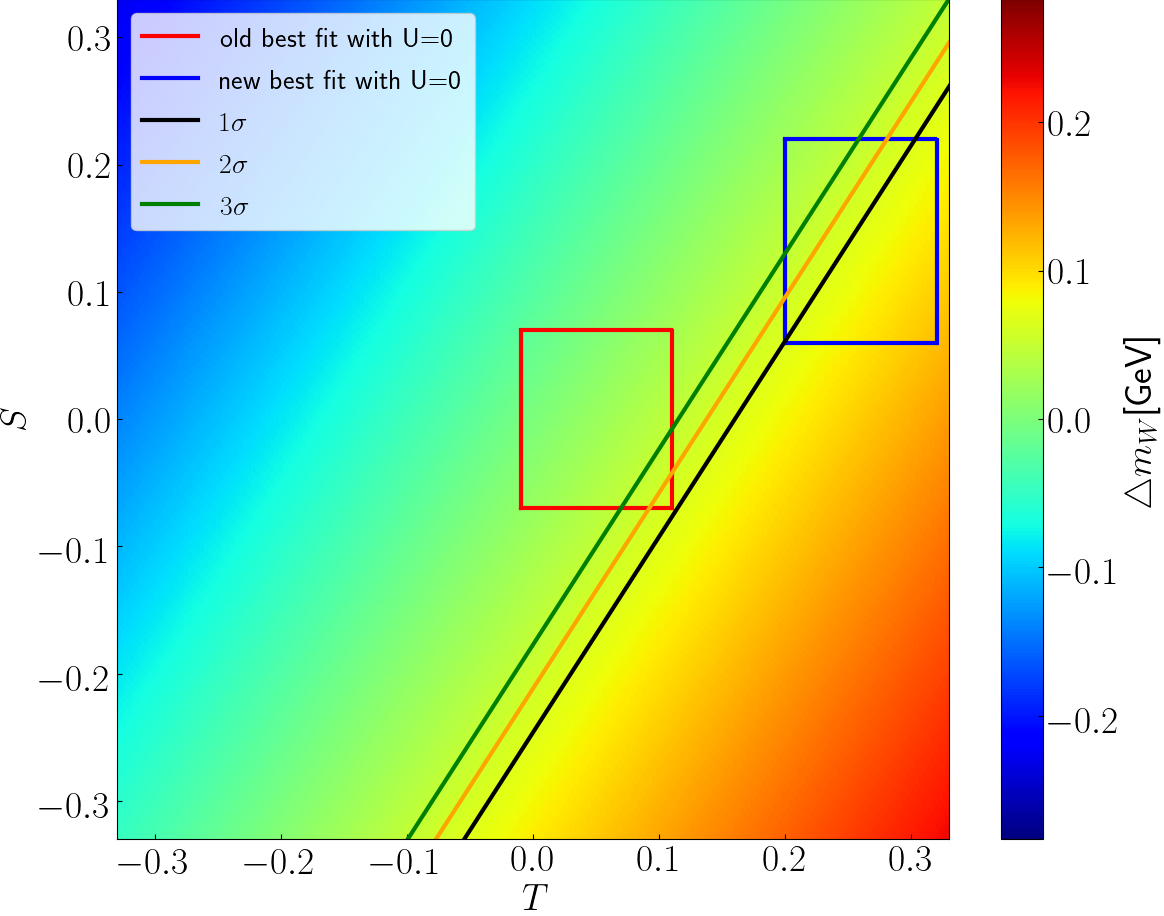}
\end{center}
\vspace{-.5cm}
\caption{Possible values of $S,T$ with $U=0$ that can give $\Delta m_W$ up to $1\sigma-3\sigma$ range of new CDF-II data. The box with red line denotes the range $S=0.00\pm0.07$ and $T=0.05\pm 0.06$~\cite{PDG:2020} while the box with blue line denotes the range $S=0.14\pm0.08$ and $T=0.26\pm 0.06$~\cite{CDFII:STUFIT}.}
\label{fig1}
\end{figure}

We plot in fig.\ref{fig1} the allowed values of $S,T$ with $U=0$\footnote{Fixing $U=0$ is motivated by the fact that $U$ is suppressed by an additional factor $M^2_{new}/M^2_Z$ compared to $S$ and $T$~\cite{UEQ0}. Such a choice can greatly improves the precision on $S$ and (in particularly) $T$. The tree level $\Delta \rho$ can also be absorbed into the effective $T$ parameter.} that can explain $\Delta m_W$ up to $1\sigma-3\sigma$ range of new CDF-II data by new physics contributions. The green box denotes the old fit $1\sigma$ constraints on $S$ and $T$ from various inputs combined with $M_Z$~\cite{PDG:2020}. Recently, electroweak precision fit of the oblique parameters from the analysis of precision electroweak data, including the new CDF-II result of $m_W$, gives~\cite{CDFII:STUFIT}
\beqa
S &=& 0.06 \pm 0.10,~T = 0.11\pm 0.12~, ~~~~U = 0.13\pm 0.09~,~\\
S &=&  0.14 \pm  0.08, ~T = 0.26\pm 0.06~,~~~({\rm fix }~U=0)\nn
 \eeqa
with (and without) fix $U=0$. Both box regions allow the choices of $S,T$ parameters that can explain the CDF-II $m_W$ anomaly to $2\sigma$ range.

GM-type models are constrained by the measurements of SM quantities, such as the determination of the $Zb\bar{b}$ coupling, the measurement of the Higgs boson signal strengths, the vacuum stability bounds and the unitary bounds. For example, the couplings of SM-like Higgs $h$ to the SM fermions and weak gauge bosons $V = W, Z$ are modified in GM-type models, who is arguably the simplest custodially symmetric model whose $\kappa_V$ values can be larger than unity and would be constrained by LHC measurements~\cite{1707.04176}. The deviations of the Higgs boson couplings to weak gauge bosons etc from the SM predictions can be parameterized by the $\ka_F$, $\ka_W$ and $\ka_Z$ parameters
\beqa
g_{h\bar{f}f}=\ka_F g_{h\bar{f}f}^{SM}~,~~~~g_{hVV}=\ka_V g_{hVV}^{SM}~~(V = W, Z)~~.
\eeqa
Note that $\ka_W$ and $\ka_Z$, which should take different values in the custodial symmetry breaking case with misaligned triplet VEVs, will turn to same values in the limit where the custodial symmetry is restored.
The updated LHC constraints on $\ka_V,\ka_F$ by  ATLAS~\cite{ATLAS:kappaWZ} and CMS~\cite{CMS:kappaWZ} collaborations are given as
\beqa
\ka_W&=&1.05\pm 0.06~,~\ka_Z=0.99\pm 0.06~,~~({\rm ATLAS})~\nn\\
\ka_W&=&1.02\pm 0.08~,~\ka_Z=1.04\pm 0.07~,~~({\rm CMS})~,
\label{LHC:kappaV}
\eeqa
Constraints from updated values of $\ka_t,\ka_b,\ka_\tau,\ka_\mu$ etc by LHC~\cite{ATLAS:kappaWZ,CMS:kappaWZ} can be similarly imposed
\beqa
\ka_t&=&0.94\pm 0.11~,~\ka_b=0.89\pm 0.11~,~\ka_\tau=0.93\pm 0.07~,~\ka_\mu=1.06^{+0.25}_{-0.30},~({\rm ATLAS})~\nn\\
\ka_t&=&1.01\pm 0.11~,~\ka_b=0.99\pm 0.16~,~\ka_\tau=0.92\pm 0.08~,~\ka_\mu=1.12\pm 0.21,~({\rm CMS})
\label{LHC:kappaF}
\eeqa

The code SARAH~\cite{SARAH} is used to generate the input model files to link the SPheno~\cite{Spheno} package, which has already implemented various updated collider constraints.  We require the predicted Higgs mass to lie between 124 GeV and 126 GeV. Bounds from the stability of the electroweak vacuum (or the bounded from below condition), the perturbative unitary condition in~\cite{GM24v2} are imposed. We also impose the most stringent upper bound for the $s_H$ parameter with $s_H\lesssim 0.2$~\cite{GM27}, although such a constraint can be relaxed in various parameter space regions. All the survived points need to pass the bounds in HiggsBounds~\cite{HiggsBounds}, HiggsSignals~\cite{HiggsSignals} as well as GMCALC~\cite{GMCal}.

\begin{figure}[htb]
\begin{center}
\includegraphics[width=2.9 in]{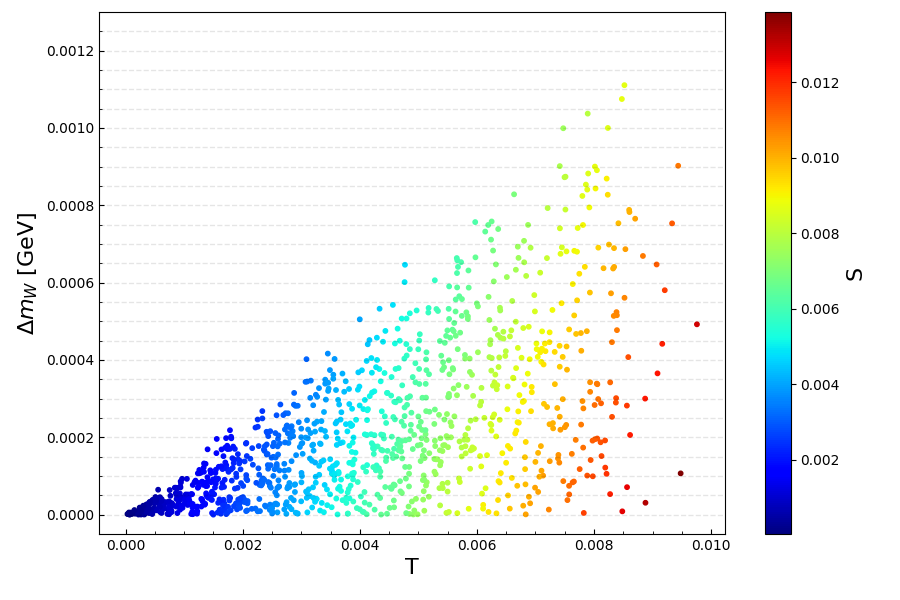}
\includegraphics[width=2.9 in]{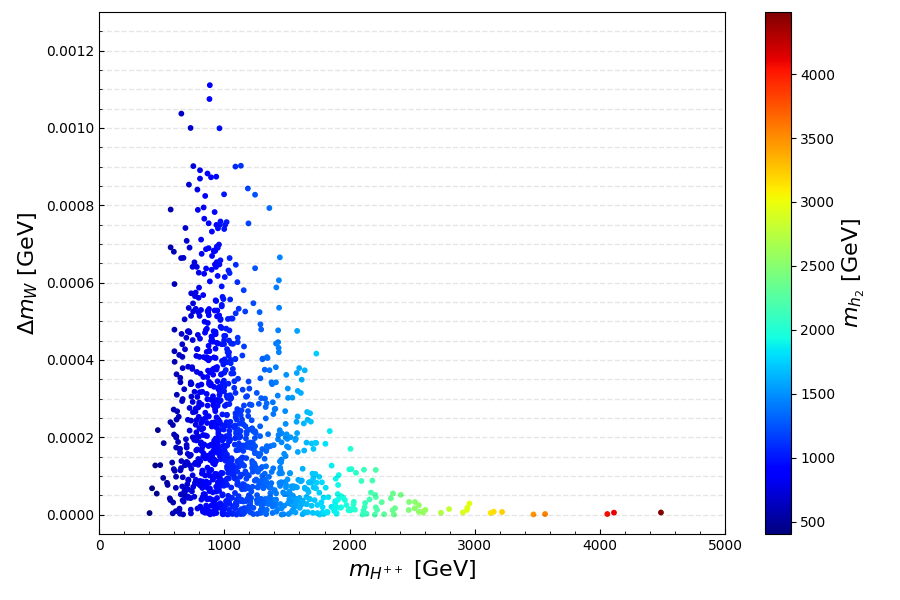}
\end{center}
\vspace{-.5cm}
\caption{We show the new physics contributions to $\Delta m_W$ in ordinary custodial symmetry preserving GM model. The dependence of $\Delta m_W$ on the mass of doubly charged scalar is also shown.}
\label{fig2}
\end{figure}

 Firstly, we scan the parameter space of ordinary GM model with aligned triplet VEVs to survey if it can successfully account for the recent W-boson mass anomaly reported by CDF-II experiment. The value of the $T$ parameter, which quantifies the strength of weak-isospin breaking through the radiative
corrections, is small in this model because of the tree-level custodial $SU(2)_c$ symmetry. So, we do not expect large new physics contributions to $m_W$ in ordinary GM model.
We show our numerical results in fig.\ref{fig2}. It can be seen from the left panel that the survived points of ordinary custodial preserving GM model can not explain the CDF-II data, which can contribute to $\Delta m_W$ a maximal amount $0.0012$ GeV at one loop level. The corresponding heavy Higgs mass $m_{h_2}$ and the doubly charged Higgs $m_{H^{++}}$ mass are also shown in the right panel with $m_{H^{++}}$ lying near 900 GeV for $\Delta m_W\sim 0.0012$ GeV. Such a heavy $H^{++}$ can give very small contributions to the muon anomalous magnetic moment $\Delta a_\mu$ by the Barr-Zee diagrams, which can not~\cite{CWChiang} account for the recently reported muon $g-2$ anomaly~\cite{g-2:FNAL}.

The $T$ parameter can be increased in GM type models if additional $SU(2)_c$ custodial symmetry breaking effects are included. The most economical recipe without putting in by hand new $SU(2)_c$ breaking terms is to split slightly the triplet VEVs, that is, requiring misalignment between $v_\xi$ and $v_\chi$ by a small amount. Such a small misalignment with real non-custodial vacuum from $SU(2)_L \tm SU(2)_R$ invariant scalar potential can still be allowed and consistent, as discussed in previous sections. Besides, it had been shown in~\cite{GM28.5} that the effects of custodial symmetry breaking by loop effects of the $U(1)_Y$ hypercharge gauge interaction in the GM model are under control up to high energy close to the theory cutoff. So we expect that the full effective scalar potential only break softly the custodial symmetry and the misalignment among the triplet VEVs should not be large. With small misalignment among the triplet VEVs, the effective $T$ parameter can be increased accordingly because of the slightly breaking of $SU(2)_c$ custodial symmetry.

\begin{figure}[htb]
\begin{center}
\includegraphics[width=2.9 in]{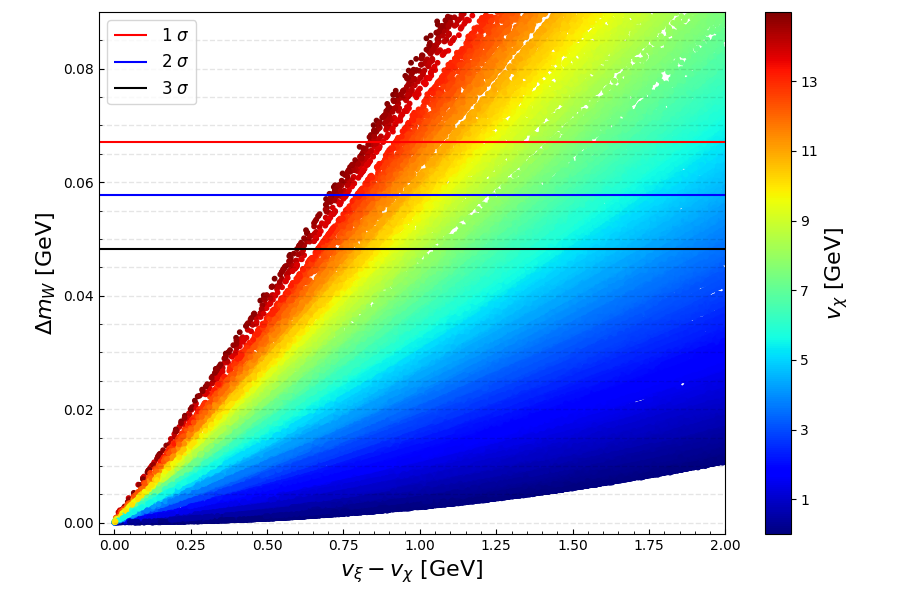}
\includegraphics[width=2.9 in]{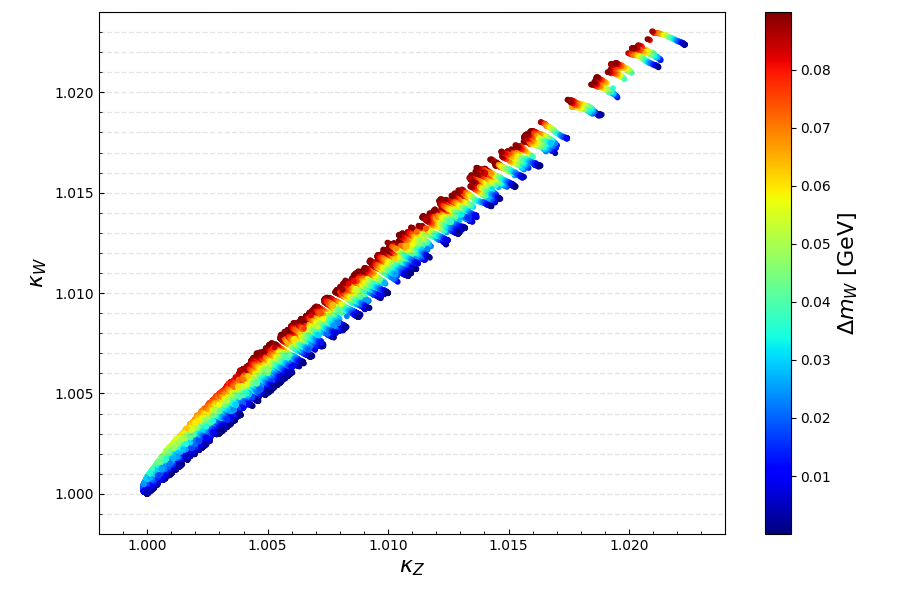}\\
\includegraphics[width=2.9 in]{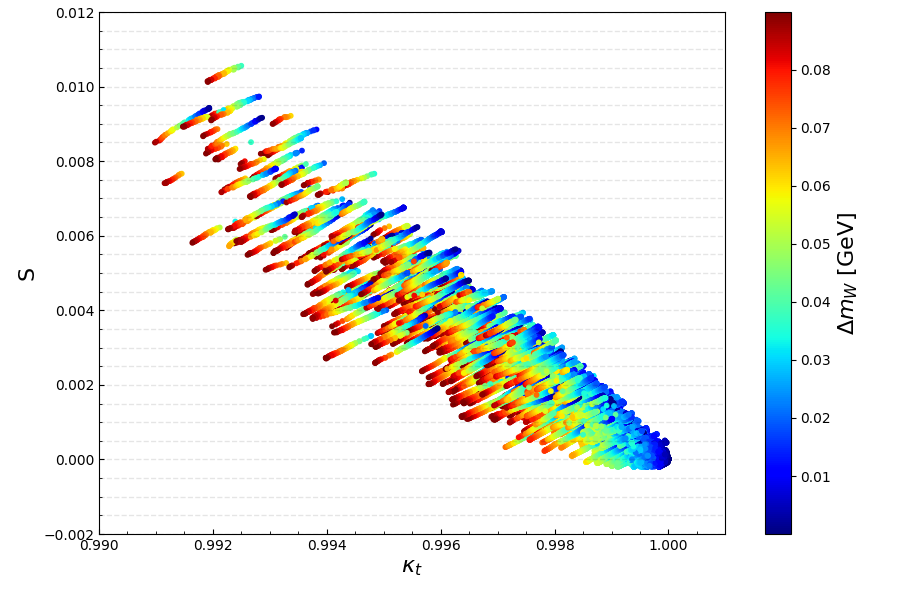}
\includegraphics[width=2.9 in]{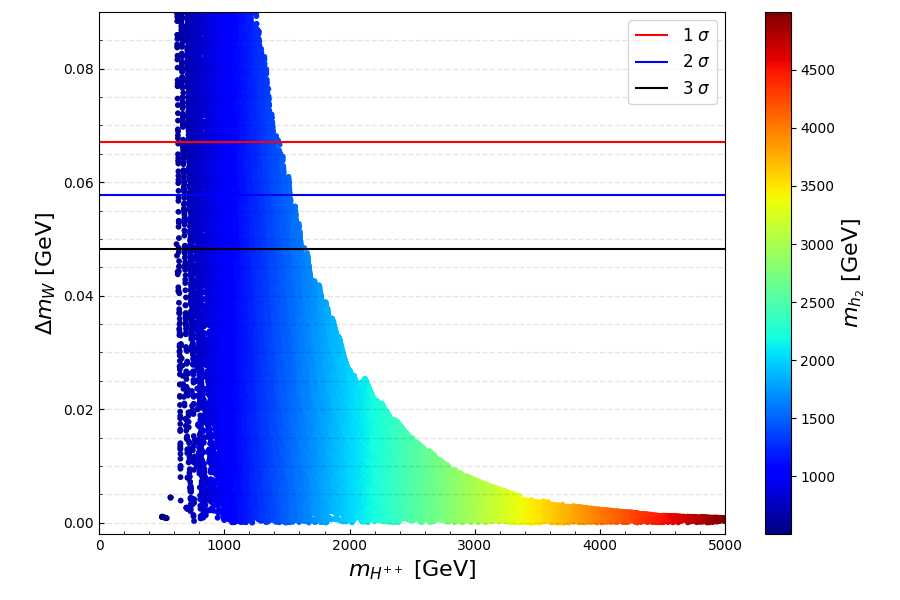}
\end{center}
\vspace{-.5cm}
\caption{We show the new physics contribution $\Delta m_W$ in the GM model with a small misalignment among the triplet VEVs. The dependences of $\Delta m_W$ on the mass of doubly charged scalar, the values of $\ka_W,\ka_Z,\ka_t$ and the $S$ parameter are also shown in the panels.}
\label{fig3}
\end{figure}
We scan the parameter space of GM model with slightly misaligned triplet VEVs and show our numerical results in fig.\ref{fig3}. From our numerical results, we can see in the upper left panel that the new physics contribution $\Delta m_W$ is indeed correlated to the $SU(2)_c$ breaking parameter $\Delta v\equiv v_\xi-v_\chi$.  New contribution to $\Delta m_W$ can easily reach $0.067$ {\rm GeV} for $\Delta v$ as small as $0.80$ {\rm GeV} with $v_\chi\lesssim 15$ GeV, explaining the CDF-II data to $1\sigma$ range. As expected, larger $\Delta v$ can increase further the value of $\Delta m_W$, which can be welcome to explain the CDF-II anomaly. Besides, for fixed $\Delta v$, larger value of $v_\chi$ always lead to larger new physics contribution to $\Delta m_W$. On the other hand, to explain the CDF-II $m_W $ data to $1\sigma$ range, larger value of $v_\chi$ needs smaller misalignment (so smaller value of $\Delta v$), which is also welcome to keep the stability of such vacuum with misaligned triplet VEVs.
From our numerical results, we can see that the values of $\ka_Z$ versus $\ka_W$ are constrained to be larger than unity (see the upper right panel of fig.\ref{fig3}), which can be seen to be consistent with the updated LHC constraints in~(\ref{LHC:kappaV}). Besides, the allowed values of $\ka_t$ (in the $1\sigma$ range of $\Delta m_W$) should lie within the narrow range $0.990\lesssim \ka_t\lesssim 0.998$, which is also compatible with the LHC constraints in (\ref{LHC:kappaF}). It can be seen from the lower right panel of fig.\ref{fig3} that the mass of $H^{++}$ Higgs is constrained to be larger than 600 GeV. Although the Barr-Zee type diagrams involving the $H^{++}$ Higgs can contribute to $\Delta a_\mu$, such a heavy $H^{++}$ (larger than almost 400 GeV~\cite{CWChiang}) can not lead to large enough $\Delta a_\mu$ to account for the new muon $g-2$ anomaly.

\section{\label{sec-4} RH-neutrino Extended GM Model}
Unlike Higgs Triplet Model (HTM) model, which extends the SM with one $SU(2)_L$ triplet, ordinary GM model always predict $\rho_{tree}=1$ by the custodial $SU(2)_c$ symmetry. So, it is fairly interesting to seek alternative custodial $SU(2)_c$ breaking sources to explain the CDF-II $m_W$ anomaly. As Yukawa couplings always break the custodial $SU(2)_c$ symmetry that preserved by the Higgs potential, we could introduce additional large $SU(2)_c$ breaking Yukawa coupling terms involving neutrinos. In ordinary GM model, the Yukawa couplings terms $h_{ij}\overline{L_L^{ic}}i\tau_2\chi L_L^j$ are responsible for neutrino masses generation, whose coupling strength should be rather tiny for $v_\Delta\sim {\cal O}(10)$ GeV. So we need to find ways to consistently increase the coupling strength $h_{ij}$. We find that large $h_{ij}$ can be naturally allowed when an additional RH neutrino sector is introduced.
 The new RH neutrino sector in our extended GM model can be written as
\beqa
-{\cal L}\supseteq y_{ij}^N \bar{L}_{L,i} \phi N_{R,j}+ \f{1}{2}(M_R)_{ij} N_{R,i}^T C N_{R,j}+h_{ij}\overline{L_L^{ic}}i\tau_2\chi L_L^j+\text{h.c.}~,
\eeqa
with $N_{R,i}$ the RH neutrinos.
After EWSB, it will lead to the neutrino mixing mass matrix of the form
\beqa
{\cal M}_\nu=\(\bea{cc} h_{ij}v_\Delta & (y_{ij}^{N})^T v_\phi \\ y_{ij}^N v_\phi & (M_R)_{ij}\eea\).
\eeqa
For simply, we choose $(M_R)_{ij}=M_{R,i}\delta_{ij}$ up to possible phases. So, the neutrino mass can be given as
\beqa
m_\nu\approx h_{ij}v_\Delta-v_\phi^2 (y_{ij}^N)^TM_{R,j}^{-1}(y_{jk}^N) ~,
\label{neutrino:cancelation}
\eeqa
for $M_R\gg v_\phi \gg v_\Delta$, which is in fact a mixed type I+II like neutrino seesaw mechanism.
Tiny neutrino mass of order $10^{-3}$ eV requires the cancelation among the two terms
\beqa
h_{ij}v_\Delta\approx \f{[(y^{N})^Ty^N]_{ij}v_\phi^2}{M_{R,i}}.
\eeqa
So, each term should take the following form
\beqas
M_{ij}^{(I,II)}=V^{*}_{PMNS} {\rm diag}(m_1,m_2 e^{-i\phi_1},m_3e^{-i\phi_1})_{ij} V_{PMNS}^{-1}~,
\eeqas
with $m_i\simeq v^2/M_{R,i}$ and $V_{PMNS}$ the Pontecorvo-Maki-Nakagawa-Sakata (PMNS) matrix ~\cite{Esteban:2020cvm}, to guarantee that the form of neutrino mass matrix can be kept after large fine-tuned cancelations.
 For $y^N\sim {\cal O}(1)$, $M_{R,i}\sim {\cal O}(1)$ TeV and $h_{ij}\sim {\cal O}(1)$, tiny neutrino mass requires $v_\Delta \sim {\cal O}(10)$ GeV, which is naturally allowed by GM model.

From previous discussions, it can be seen that the coupling $h_{ij}$ should take the following form
\beqa
h_{ij}=2\sqrt{2} (V_{PMNS}^T)_{i m}^{-1} \(\f{v (1-s_H^2)}{s_H M_{R;m}}\)\delta_{m n}(V_{PMNS})_{n j}^{-1}~.
\label{hijcoupling}
\eeqa
 However, the coupling strength $|h_{ij}|\sim {\cal O}(1)$ will be constrained stringently by the lepton flavor violation (LFV) processes, especially when it is not very small. In fact, the main LFV signatures of the seesaw mechanism stem from muon decay, with the current bounds~\cite{LFV:PDG}
 $$
 ~BR(\mu\ra e \ga)< 4.2\tm 10^{-13},~~BR(\mu \ra 3e)< 10^{-12}.
 $$
 Other LFV process, such as $\mu-e$ conversion and LFV decays involving $\tau$ lepton, are subdominant~\cite{LFV:mue,LFV:mue2}. For scalar triplets masses of order 1 TeV, the typical magnitude of $|h_{ij}|$ is constrained to be less than ${\cal O}(10^{-2})$ by the $BR(\mu\ra e \ga)$ bound
\beqa
BR(\mu\ra e \ga)\simeq \f{\al_{EM}}{192\pi G_F^2}|(h^\da h)_{e\mu}|^2 \(\f{1}{M^2_{H^{\pm\pm}}}+\f{8}{M_{H^\pm}^2}\)^2~.
\eeqa
Due the correlation between $h_{ij}$ and $M_{R,i}$ via eq.(\ref{hijcoupling}), the scale of $M_{R,i}$ (for simply, we set $M_{R,i}\approx M_R$ ) should typically be heavier than 50 TeV for $s_H<0.2$ and $M_{H^{\pm\pm}}\simeq m_\Delta$ at 1 TeV.

We should note again that, in ordinary GM model, the coupling $h_{ij}$ should be tiny (of order $10^{-13}$) for $v_\Delta\sim {\cal O}(10)$ GeV. With the augmentation of the RH neutrino sector, the allowed value of $h_{ij}$ can be greatly increased. On the other hand, to get the observed tiny neutrino masses, the price of large fine-tuning (FT) had to be paid for the cancelation in eq.(\ref{neutrino:cancelation}). The larger the $M_R$ value, consequently a smaller $h_{ij}$ coupling term (hence a smaller $SU(2)_c$ breaking source), the smaller the FT is needed.

The introduction of RH neutrinos can be advantageous in cosmology. Although the realization of electroweak baryogenesis can be possible in GM type model after adding additional CP violation sources~\cite{GM21}, the idea that the baryon asymmetry is induced by the interactions responsible for neutrino masses (the leptogenesis) is fairly attractive. In leptogenesis, one-loop non-vanishing CP asymmetry requires flavor "breaking", or more exactly  at least two sources of flavor "breaking", i.e. two heavy states with unequal couplings to leptons and/or scalar bosons. So, pure type-II seesaw model (and also in GM model) with a single scalar triplet, which is the only
one that can give 2 or 3 light neutrino masses from a single heavy state, gives a vanishing asymmetry, even though it a priori satisfies all Sakharov conditions~\cite{Hambye:2012fh,Ma:1998dx}.
Therefore, to successfully adopt the leptogenesis mechanism, it is attractive to reintroduce heavy Majorana neutrinos, which can be responsible for both the observed baryon asymmetry in the universe and the generation of the neutrino masses via mixed type I+II like seesaw mechanism. It has been found in~\cite{hep-ph:0309342,Pilaftsis:2003gt} that, for leptogenesis in the framework of the type-I seesaw model, a lower bound of about $10^8\sim 10^9$ GeV on the RH neutrino scale can be derived if the heavy singlet neutrinos have an hierarchical mass spectrum. To obtain this lower bound, the size of the leptonic asymmetry between the heavy Majorana neutrino decay $N\ra L \phi$ and its respective CP conjugate plays a key role. It has been shown in~\cite{hep-ph:0309342} that the leptonic CP asymmetry is not only analytically well-behaved but also can be of order unity if two of the heavy Majorana neutrinos have mass differences comparable to their decay widths. Due to this resonant effect, the allowed RH neutrino mass scale from the requirement of generating sufficient lepton number asymmetry in resonant leptogenesis scenario can be
significantly lower. Even TeV scale RH neutrino masses ($M_1,M_2\sim {\cal O}(1)$ TeV) are still allowed to obtain sufficient baryon asymmetry. So, with ${\cal O}(100)$ TeV Majorana RH neutrino masses,  leptogenesis with type-I seesaw can already generate the required baryon asymmetry in the universe.
On the other hand, our RH neutrino extended GM model in fact leads to a mixed type I+II neutrino seesaw mechanism. With both RH neutrinos $N_{R,j}$ and a single scalar triplet $\chi$, the corresponding leptogenesis can work well and have new CP asymmetry sources~\cite{Leptogenesis:I+II}, such as $\epsilon_{N_k}^\chi$ from the $N_k$ decay and $\epsilon_{\chi}$ from the $\chi$ triplet decay.  Because of large cancelations between the type-I and type-II seesaw contributions to the neutrino masses, both (type-I and type-II) contributions to lepton asymmetry by leptogenesis are important~\cite{Hambye:2003ka}. In addition to the contributions to leptogenesis from the decay of the RH neutrino singlets to leptons and Higgs scalar, the contributions from the decay of the triplet to two leptons can also generate sizeable lepton asymmetry.

\begin{figure}[htb]
\begin{center}
\includegraphics[width=2.9 in]{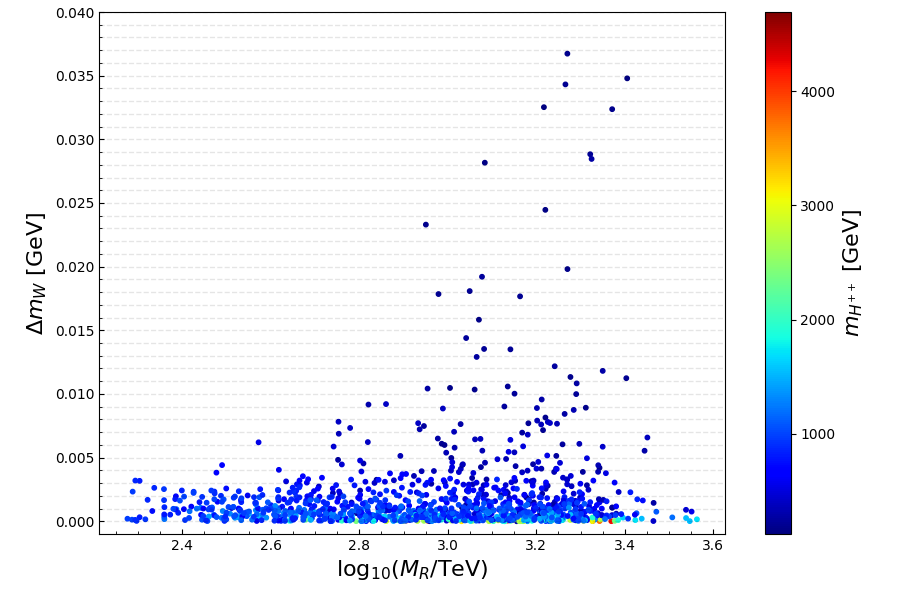}
\includegraphics[width=2.9 in]{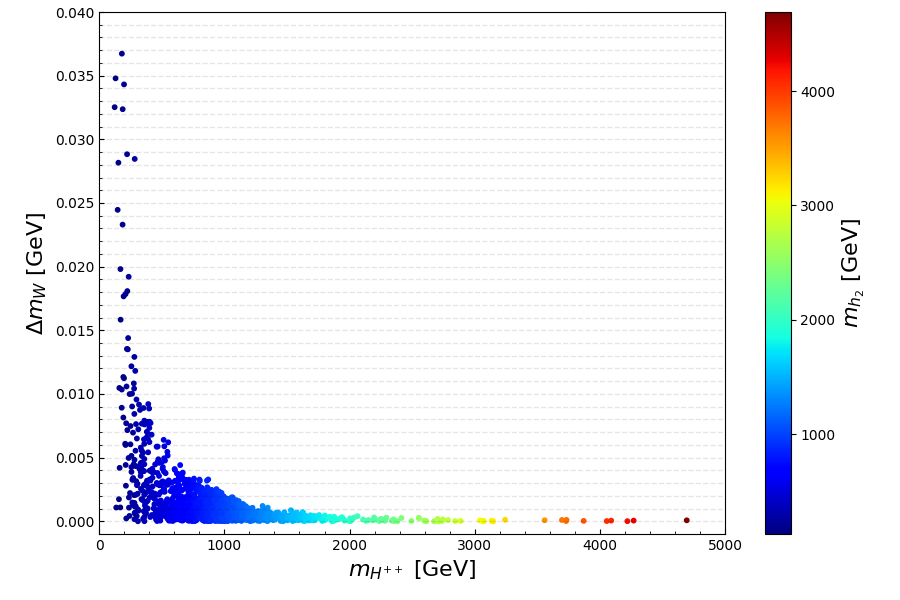}\\
\includegraphics[width=2.9 in]{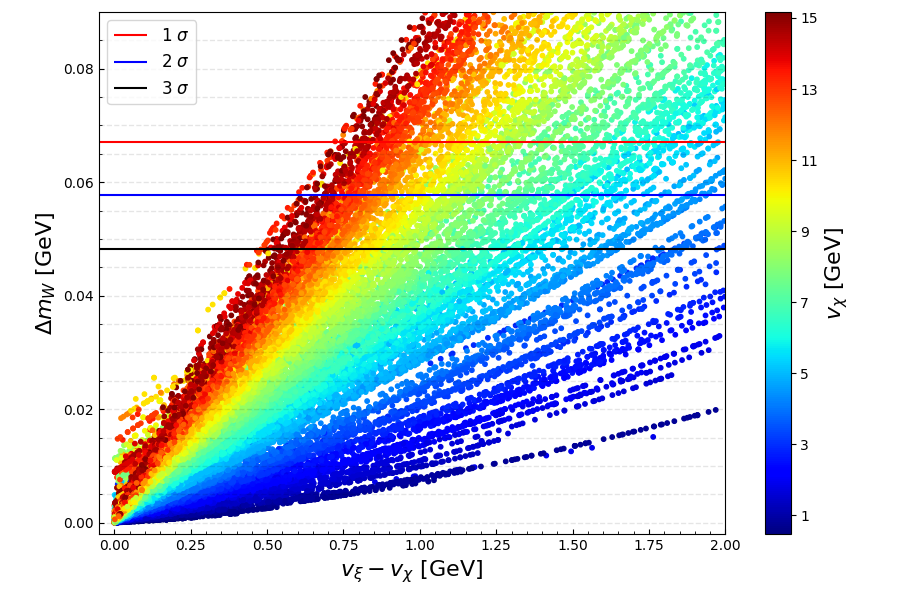}
\includegraphics[width=2.9 in]{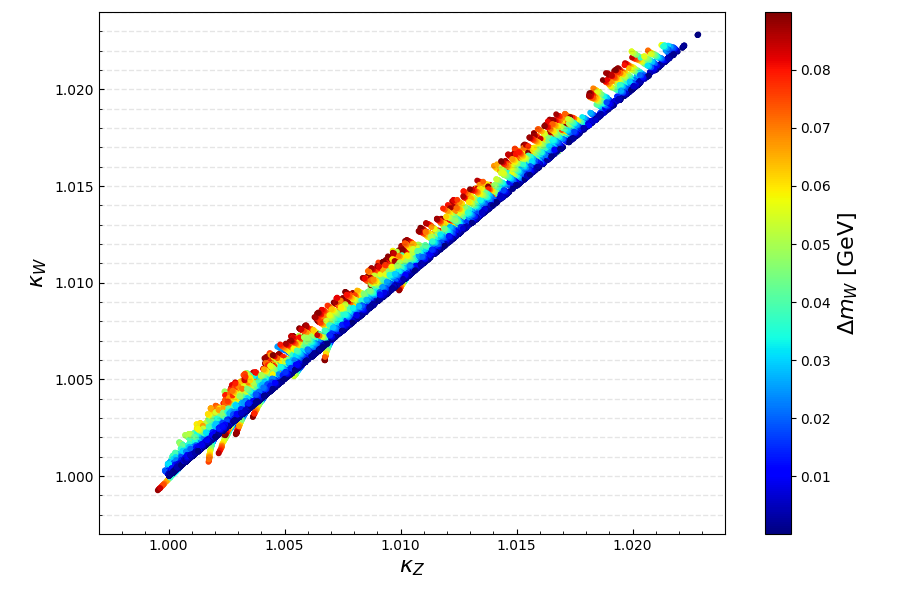}\\
\includegraphics[width=2.9 in]{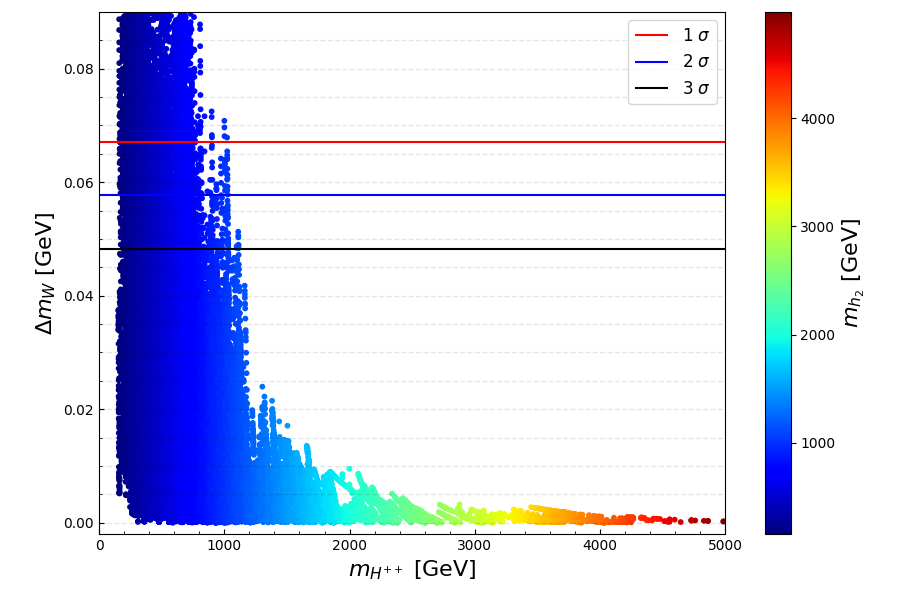}
\includegraphics[width=2.9 in]{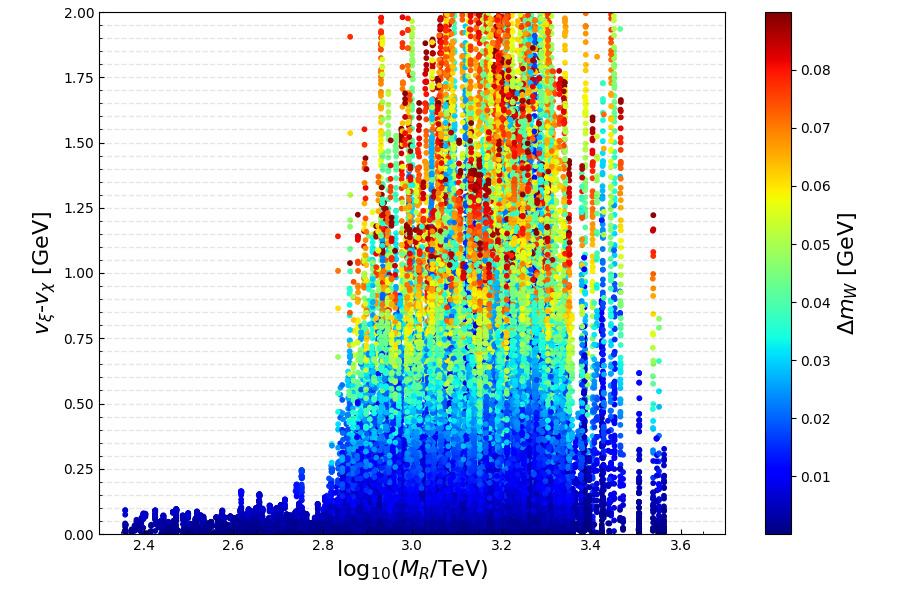}\\
\end{center}
\vspace{-.5cm}
\caption{We show the new physics contribution $\Delta m_W$ in the RH neutrino extended GM model without (the upper panels) and with (the middle and the lower panels) small misalignment among the triplet VEVs. The corresponding ranges of $\ka_W$, $\ka_Z$, $M_R$ and the doubly charged scalar masses $m_{H^{++}}$ etc are also shown.}
\label{fig4}
\end{figure}

 GM extension model with RH neutrinos can allow moderately large Yukawa coupling strength $|h_{ij}|\sim {\cal O}(10^{-2})$ with large FT in the cancelation between type I and type II type contributions for neutrino masses to generate tiny neutrino masses of order $10^{-3}$ eV. So the introduction of RH neutrinos are crucial to allow large $h_{ij}{L}_L^T \chi L_L $ coupling. Larger values of $h_{ij}$ coupling will lead to larger tree level $SU(2)_c$ symmetry breaking effects, which can be welcome to increase $\Delta m_W$ by loop contributions. We show the contributions to $\Delta m_W$ with moderately large $h_{ij}$ terms in the upper panels of fig.\ref{fig4} for our RH neutrino extended GM model. We can see from the upper left panel of fig.\ref{fig4} that, although the new physics contribution $\Delta m_W$ in the RH neutrino extended GM model is much larger than that in the ordinary custodial symmetry preserving GM model (with aligned triplet VEVs), this RH neutrino extension model without misalignment among the triplet VEVs can still not explain the new CDF-II anomaly on W-boson mass. Lower scale of $M_R$, which corresponds to larger value of $h_{ij}$, can possibly increase the value of $\Delta m_W$, with the maximal contribution about $0.03$ GeV. However, most of such points are ruled out by LFV constraints. Therefore, large $h_{ij}$ term with the introduction of the RH neutrino sector can greatly increase the new physics contributions to $\Delta m_W$ in comparison to ordinary GM model, but it is still insufficient to explain the new CDF-II $m_W$ anomaly.

Although merely introducing small misalignment among the triplet VEVs can already explain the CDF-II $m_W$ anomaly up to $1\sigma$ range, it is still interesting to see how the presence of both custodial breaking sources can affect the degree of misalignment needed to explain the anomaly.
That is, we would like to combine both the effects of moderately large $h_{ij}$ coupling and small misalignment among the triplet VEVs. As discussed previously, real custodial symmetry breaking vacuum can be the true/metastable minimum of the $SU(2)_L\tm SU(2)_R$ invariant scalar potential. The pseudo-Goldstone modes $H_5^\pm\sim (\chi^\pm-\xi^\pm)/\sqrt{2}$, which correspond to the breaking of custodial $SU(2)_c$ to residual $U(1)$, can receive large radiative contributions to their masses even for a TeV UV-completion scale, as the $h_{ij}$ coupling can be large after the introduction of RH neutrinos.

 It can be seen in the middle and lower panels of fig.\ref{fig4} that large new physics contributions to $\Delta m_W$ can still be easily given in this RH neutrino extended GM model (accompanied with small misalignment among the triplet VEVs). Besides, the splitting needed to explain the CDF-II $m_W$ anomaly up to $1 \sigma$ range is smaller than that without RH neutrino sector. For example, from the middle left panel, it is obvious that to explain the CDF-II W boson mass anomaly to $1\sigma$ range, the minimal splitting among the triplet VEVs should be larger than $0.70$ GeV for $v_\chi\lesssim 15$ GeV.  The corresponding values of $\ka_W,\ka_Z$ and the $m_{H^{++}}$, $m_{h_2}$ masses are also shown in these panels, which lie within the LHC bounds. The $1\sigma$ ranges of CDF-II $m_W$ data require $\ka_W,\ka_Z$ to be larger than unity and smaller than 1.022 (and 1.020), respectively. The mass range for $m_{H^{++}}$ allows lighter (than 400 GeV) $H^{++}$, which can possibly explain the muon $g-2$ anomaly to $3 \sigma$ range by Barr-Zee type contributions. The combination of both custodial symmetry breaking effects can give constructive contributions to $\Delta m_W$. The loose correlations between the RH neutrino sector and the needed splitting of the triplet VEVs are shown in the $M_R$ versus $\Delta v$ plot (the lower right panel).



\section{\label{sec-5} Conclusions}
   GM model, which extends the SM with new triplets and preserves the tree level custodial symmetry, can hardly account for the new CDF-II anomaly on W-boson mass in its original form.  As expected, unless additional $SU(2)_c$ custodial symmetry breaking effects are significant, the new physics contributions to $\Delta m_W$ in GM type model are always very small. Our numerical results show that ordinary GM model can contribute to $\Delta m_W$ a maximal amount $0.0012$ GeV, which can not explain the new CDF-II data on W boson mass.

   We propose firstly to introduce small misalignment among the triplet VEVs in GM model to break the custodial $SU(2)_c$ symmetry so as that the new physics contributions to $\Delta m_W$ can be increased.  Such slightly misaligned triplet VEVs from tree level custodial symmetry preserving scalar potential (appended with custodial symmetry breaking loop level effective potential) can still be allowed, because no disruptive jumping transition from the true stable vacuum in the aligned case to an unstable saddle point vacuum in the misaligned case should occur when one begins to slightly turn on small misalignment among the triplet VEVs (unless the splitting within the misaligned triplet VEVs is big enough) from a custodial preserving tree level scalar potential (with custodial symmetry breaking loop level contributions) by continuously varying the relevant parameters. Our numerical results indicate that the resulting $\Delta m_W$ can easily reach the $1\sigma$ range of CDF-II $m_W$ data and the splitting among the triplet VEVs $\Delta v=v_\xi-v_\chi$ can be as small as $0.8$ GeV for $v_\chi\lesssim 15$ GeV.

  In ordinary GM model, the neutrino masses can be generated by the $h_{ij}$ term via the type II seesaw like mechanism with a very tiny $h_{ij}$ coupling strength for triplet VEVs of order GeV. We also propose to extend the GM model with a low scale RH neutrino sector, which can be used to generate tiny neutrino masses by a mixed type I+II like neutrino seesaw mechanism and realize successfully leptogenesis mechanism, as type-II leptogenesis with a single triplet alone does not work. Given the tiny neutrino masses, the $h_{ij}$ coupling strength can be much larger (even of order $10^{-2}$) in this case. Larger $h_{ij}$ term means a larger $SU(2)_c$ breaking source. With low scale RH neutrino mass scale of order $10^2\sim 10^4$ TeV, the new physics contribution to $\Delta m_W$ can reach $0.03$ GeV and is much larger than that of ordinary custodial symmetry preserving GM model. However, such a $\Delta m_W$ value is insufficient to explain the CDF-II $m_W$ anomaly

 It is interesting to combine both $SU(2)_c$ breaking effects, the small misalignment among the triplet VEVs and moderately large $h_{ij}$ couplings. Moderately large $h_{ij}$ term, allowed by the extension with a RH neutrino sector, can give large contributions to the masses of the pseudo-Goldstone modes $H_5^\pm$ even with low UV completion scale. As moderately large $h_{ij}$ term acts as an additional $SU(2)_c$ breaking source other than the $U(1)_Y$ hypercharge (and Yukawa) interactions, we expect that the full scalar potential including the custodial symmetry breaking loop corrections should easily lead to a custodial symmetry breaking minimum with small misalignment among the triplet VEVs. In this case, our numerical results show that the value of $\Delta m_W$ can still easily reach the $1\sigma$ range of CDF-II $m_W$ data, with a minimum splitting (among the triplet VEVs) approximately $0.7$ GeV for $v_\chi\lesssim 15$ GeV.

\begin{acknowledgments}
We are very grateful to the editor and referee for helpful suggestions. This work was supported by the National Natural Science Foundation of China (NNSFC) under grant Nos. 12075213, by the Key Research Project of Henan Education Department for colleges and universities under grant number 21A140025, by the National Supercomputing Center in ZhengZhou.
\end{acknowledgments}

\end{document}